\documentclass[aps,prl,twocolumn,superscriptaddress]{revtex4-2}  
\usepackage{amsfonts}
\usepackage{bbding}
\usepackage{amssymb}
\usepackage[title]{appendix}
\usepackage{color}
\usepackage{amsmath}
\usepackage{graphicx}
\usepackage{subfigure}
\usepackage{txfonts}
\usepackage{soul}
\usepackage{bm}
\usepackage[normalem]{ulem}
\usepackage{natbib}
\usepackage{calrsfs}
\usepackage{tikz}
\usetikzlibrary{graphs, positioning, quotes} 
\usetikzlibrary{matrix}
\usepackage[colorlinks,linkcolor=blue,citecolor=blue,urlcolor=blue]{hyperref}
\usepackage{orcidlink}
\preprint{APS/123-QED}
\usepackage{xcolor}
\usepackage{natbib}

\DeclareMathAlphabet{\pazocal}{OMS}{zplm}{m}{n}

\newcommand{\Ds}{\pazocal{D}}
\newcommand{\Cs}{\pazocal{C}}
\newcommand{\SSE}{stochastic Schr\"odinger equation}

\begin{document}

\title{ Signatures of Environment-Induced Quantum Synchronization Transitions via Two-body Dissipator Engineering}

\author{Xingli Li\orcidlink{0000-0003-2339-6557}}
\affiliation{Department of Physics, The Chinese University of Hong Kong, Shatin, New Territories, Hong Kong, China}

\author{Yan Li\orcidlink{0009-0007-7218-133X}}
\affiliation{Department of Physics, The Chinese University of Hong Kong, Shatin, New Territories, Hong Kong, China}

\author{Yangqian Yan\orcidlink{0000-0002-3237-5945}}
\email{yqyan@cuhk.edu.hk}
\affiliation{Department of Physics, The Chinese University of Hong Kong, Shatin, New Territories, Hong Kong, China}
\affiliation{State Key Laboratory of Quantum Information Technologies and Materials, The Chinese University of Hong Kong, Hong Kong SAR, China}
\affiliation{The Chinese University of Hong Kong Shenzhen Research Institute, 518057 Shenzhen, China
}

\begin{abstract}
Metronome synchronization and the transition between the in-phase and anti-phase synchronization have been observed in classical systems. We demonstrate the quantum analog of this phenomenon in a two-qubit system coupled to a common environment.
Tracing out the environment in the quantum collision model, we obtain an effective master equation with a two-body dissipator for two qubits.
Quenching  the two-body dissipator, we demonstrate controlled transitions from in-phase to anti-phase synchronization.
This synchronization transition is robust against noise.
Signatures of the transition are observed through Pearson correlation coefficient measurements obtained via quantum simulations on superconducting circuits.
Future experiments employing qutrit systems are expected to yield a more pronounced effect.

\end{abstract}
\date{\today}

\maketitle

\begin{figure}
    \includegraphics[width=0.95\linewidth]{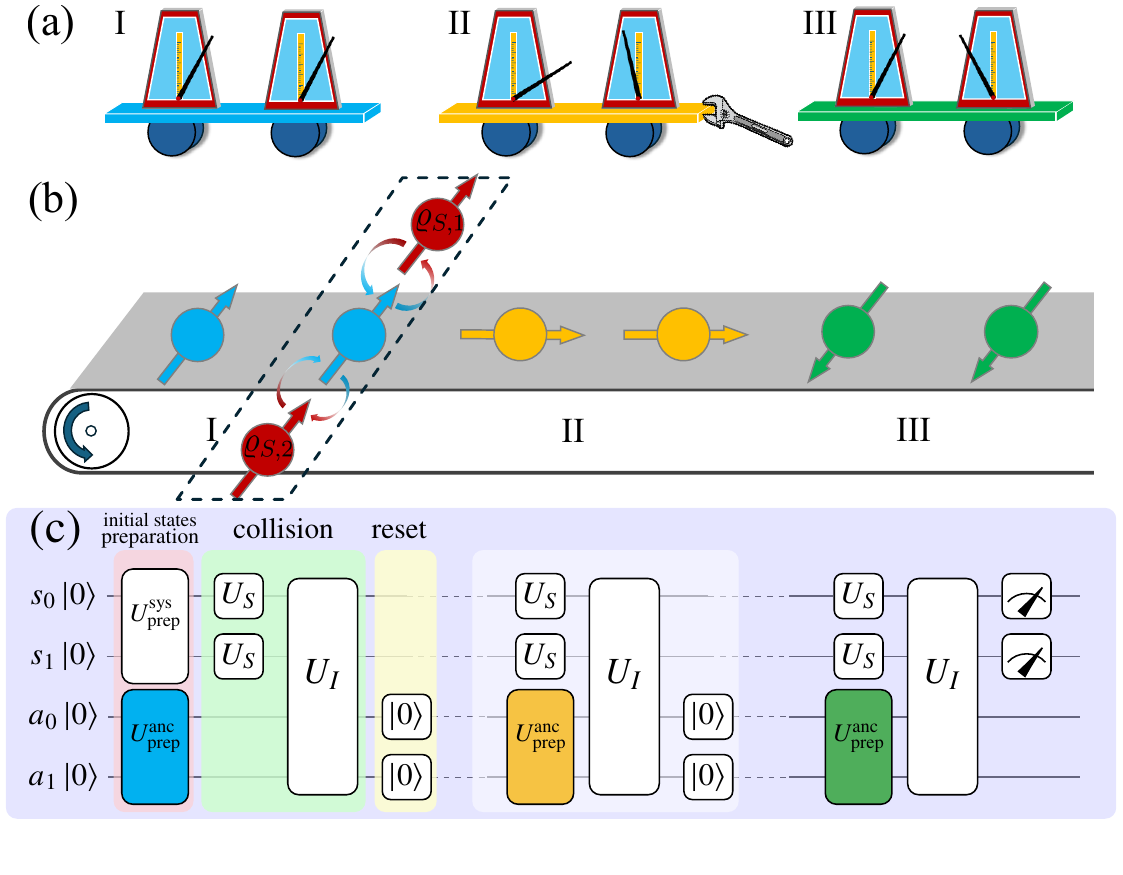}
    \caption{\label{Fig:Sketch}
    (a) Schematics of classical synchronization transition with two metronomes coupled by a common movable platform. The transition from in-phase (I) to anti-phase (III) synchronization can be realized by tuning the movable platform (II).
    (b) Schematics of the quantum collision model.
    The two qubits $\varrho_{S,1(2)}$ interact stroboscopically with ancilla qutrits (environment).
    Different ancilla states lead to different effective dissipators for the qubits. Quantum synchronization transition is induced by quenching the ancilla state. 
    (c) Schematics of the quantum circuit that simulates the quantum collision model.
}    
\end{figure}

{\it Introduction.--} 
In-phase and anti-phase synchronization have been observed in classical metronome systems placed on movable platforms.
It has also been demonstrated that the transition between these synchronization states can be achieved by engineering the common environment [Fig.~\ref{Fig:Sketch}(a)], i.e., by tuning the properties of the shared platform~\cite{kapitaniak2012synchronization,kapitaniak2014synchronous}.
To date, research on synchronization has expanded into the quantum domain, with numerous studies exploring synchronization in a wide range of open quantum systems~\cite{MariPRL2012, GiorgiPRA2013, BellomoPRA2017, Buca2019Apr, Buca2022Mar, LiPRA2023, NadolnyPRL2023, Vaidya2024Mar,Solanki2024b,Cabot2019,Shen2023b,Wachtler2023a,Murtadho2023a,Solanki2023a,Cabot2021May}, like van der Pol oscillators~\cite{Lee2014Feb, Walter2014Mar,Aifer2024} and spin chains~\cite{SchmolkePRL2022, SchmolkePRL2024, Wachtler2024May}.
Synchronization has also been demonstrated across diverse platforms such as cold atoms~\cite{Laskar2020Jul}, trapped ions~\cite{Zhang2023Sep,Li2025d}, nuclear spins~\cite{KrithikaPRA2022}, and superconducting circuits~\cite{Koppenhofer2020Apr}.

While the classical synchronization transitions can be induced purely by modifying the common environment without altering the coupling form between the metronomes and the movable platform, the quantum version of transitions between in-phase and anti-phase synchronization typically relies on adjusting the inter-system couplings~\cite{Ying2014Nov, Zhou2023Oct}.
This highlights a fundamental gap: a quantum framework for environmental engineering, in which dissipation is tuned to control synchronization, remains underdeveloped. 
To bridge this gap, we introduce a pathway for dissipation engineering: controlling the dissipation by designing the states of the environment, while leaving the system-environment interaction Hamiltonian unchanged.

To achieve it in a quantum system, we utilize the quantum collision model (QCM)~\cite{RauPR1963, CavesPRA1987, ScaraniPRL2002, ZimanPRA2002, KarevskiPRL2009, GiovannettiPRL2012, CiccarelloPRA2013, CattaneoPRL2021,CattaneoPRXQ2023,Karpat2021,Zhang2023Apr}, which is also known as the repeated-interaction scheme [Fig.~\ref{Fig:Sketch}(b), see Supplementary Materials (SM)~\cite{Supple,BreuerBook2002,Fidelity,SSEbook, Wiseman2009Nov,BrownSHPSB2009,CiccarelloPR2022,EffectiveOQSReiter2012Mar} for details].
The QCM provides a versatile platform for engineering the environment, as it allows for precise control over the states of individual ancillas and their interactions with the system.
Furthermore, since the interaction between the ancillas and system is unitary, it can be easily realized on quantum circuits~\cite{CattaneoPRXQ2023}.
Within our model, we introduce a ``quantum analog" of the classical metronome synchronization experiment: two noninteracting qubits (``quantum metronomes") couple to the common environment (``movable platform"), which may lead to dissipative coupling, as shown in Fig.~\ref{Fig:Sketch}(a).
We have the following findings: 
(i) Synchronization, which is robust to noise, spontaneously emerges in quantum systems once a weak coupling is established through a common environment.
(ii) Tracing out the environment, we obtain an effective master equation with its two-body dissipator determined by the state of the environment. While one-body dissipator and dark state engineering have been extensively studied~\cite{Murch2012,Morigi2015,Ma2019,Lin2022b}, to the best of our knowledge, two-body dissipator engineering has not been previously well explored.
(iii) We demonstrate the transition of in-phase and anti-phase synchronization via quenching the environment, or effectively, the two-body dissipator.
(iv) We observe the signatures of the transition from the Pearson correlation coefficient by performing quantum simulation on superconducting qubits.

{\it Model.---} 
The Hamiltonian of ``metronomes" reads $H_{S}=\sum_{j=1,2}\hbar\omega\sigma^{z}_{j}$, where $\sigma^z_{j}$ is the Pauli matrix and $\omega$ is half the transition frequency. We model the environment as a stream of three-level ancillas (qutrit), labeled as $\{|g\rangle,|e\rangle,|r\rangle\}$.
Each ancilla sequentially interacts with ``metronomes" [Fig.~\ref{Fig:Sketch}(b)], governed by the interaction Hamiltonian: $H_{I}=\hbar g(\sigma_{1}^{-}\otimes |r\rangle\langle g|+\sigma_{2}^{-}\otimes |r\rangle\langle e|+\text{H.c.})$, where $g$ is the coupling strength, $\sigma^{-}_{j}$ is the lowering operator for the $j$-th qubit.
The evolution operator $U$  during a collision period, consists of two contributions: $U=U_{I}U_{S}=e^{-iH_{I}\tau/\hbar}e^{-iH_{S}\tau/\hbar}$, where $\tau$ is the time for each collision. 
After interacting with the ancilla state $\eta^{n+1}_{E}$ at the $(n+1)$-th collision, the density matrix of the system $\varrho^{n}_{S}$ evolve according to a quantum map $\Phi$,
\begin{equation}
    \varrho^{n+1}_{S}=\Phi[\varrho^{n}_{S}]:=\text{tr}_{\eta^{n+1}_{E}}[U\varrho^{n}_{S}\otimes\eta^{n+1}_{E}U^{\dagger}].
\end{equation}
Altering ancilla states enables us to engineer the properties of the environment or the quantum analog of ``movable platform". 
Under Markov approximation, we obtain that, for any given ancilla state $\eta_{E}=|\psi_{E}\rangle\langle\psi_{E}|$, where $|\psi_{E}\rangle=\cos{\theta}|g\rangle+\sin{\theta}e^{i\phi}|e\rangle$, QCM predicts the effective master equation, 
(see SM~\cite{Supple} for detailed derivation)
\begin{equation}
\frac{d}{dt}\varrho_{S}=-\frac{i}{\hbar}[H_{S},\varrho_{S}]+\frac{g^2\tau}{\hbar^2}
\Ds_{\varrho_{S}}[\cos{\theta}\sigma^{-}_{1}+\sin{\theta}e^{i\phi}\sigma^{-}_{2}]
\label{Eq:ME}
\end{equation}
where $\Ds_{\varrho_{S}}[o] = o\varrho_{S}o^{\dagger}-\frac{1}{2}(o^{\dagger}o\varrho_{S}+\varrho_{S}o^{\dagger}o)$ is the effective dissipator. 
We consider three distinct dynamical phases: for $|\psi_{E}\rangle_{\text{I}}$ in {\it Phase} I, $\theta=\pi/4$, $\phi=\pi$; for $|\psi_{E}\rangle_{\text{II}}$ in {\it Phase} II,  $\theta=0$; and for $|\psi_{E}\rangle_{\text{III}}$ in {\it Phase} III, $\theta=\pi/4$, $\phi=0$.
The above master equation can be viewed as the system interacting with a bath of multiple ancillas~\cite{RodriguesPRL2019}, or as the system interacting with an ancilla, yet the ancilla itself is undergoing significant spontaneous emission (see also SM~\cite{Supple}).

Before we discuss the synchronization transition, we explain the synchronization mechanism within our system through Eq.~(\ref{Eq:ME}): two-body dissipation gives rise to dark states and decoherence-free subspace spanned by dark states~\cite{Lee2014Feb}.
A dark state $|\psi_{D,j}\rangle$ is defined as a state that is immune to dissipative process, i.e., $o|\psi_{D,j}\rangle=0$, with $H_{S}|\psi_{D,j}\rangle=\epsilon_j|\psi_{D,j}\rangle$. Thus, the corresponding eigenoperators $A_{jk}=|\psi_{D,j}\rangle\langle\psi_{D,k}|$ satisfy $\Ds_{A_{jk}}[o]=0$. For instance, in the case of $\Ds_{\varrho_{S}}[\frac{\sigma^{-}_{1}-\sigma^{-}_{2}}{\sqrt{2}}]$, which corresponds to {\it Phase} I, two dark states emerge: the ground state $|\psi_{D,1}\rangle=|\downarrow\downarrow\rangle$ and the symmetric state $|\psi_{D,2}\rangle=\frac{|\downarrow\uparrow\rangle+|\uparrow\downarrow\rangle}{\sqrt{2}}$, where $|\downarrow\rangle$ and $|\uparrow\rangle$ denote the eigenstates of $\sigma^{z}$.
In the long-time limit, the steady-state density matrix reads
$\varrho^{\text{ss}}_{S}\sim \sum_{j,k=1,2}c_{jk}e^{i(\varepsilon_{k}-\varepsilon_{j})n\tau}A_{jk}$, the constants $c_{jk}$ are determined by the initial state, satisfying $c_{11} + 2c_{22}=1$ to preserve trace.
In this study, we employ the $x$-direction magnetizations as the order parameter, denoted by $\langle \sigma^{x}_{j}\rangle_{n} = \text{tr}[\varrho^{n}_{S}\sigma^{x}_{j}]$, to signify the phase relationship between the two spins.
Note that when $c_{11}$ or $c_{22}$ equals 0, the final density matrix is a pure state and there exists no oscillation in the order parameter; for a pure state, the oscillation reaches its maximum at $c_{11}/c_{22}=2$.
Likewise, for $\Ds_{\varrho_{S}}[\sigma^{-}_{1}]$, dark states are $|\psi_{D,1}\rangle=|\downarrow\downarrow\rangle$ and $|\psi_{D,2}'\rangle=|\downarrow\uparrow\rangle$, and  
for $\Ds_{\varrho_{S}}[\frac{\sigma^{-}_{1}+\sigma^{-}_{2}}{\sqrt{2}}]$, dark states are the ground state $|\psi_{D,1}\rangle=|\downarrow\downarrow\rangle$ and the antisymmetric state $|\psi_{D,2}''\rangle=\frac{|\downarrow\uparrow\rangle-|\uparrow\downarrow\rangle}{\sqrt{2}}$. 
Through the decomposition of steady state, we reexpressed the order parameters as $\langle\sigma^{x}_{j}\rangle\sim \text{tr}[(A_{12} + A_{21})\sigma^{x}_{j}]$. 
Therefore, the phase difference between them is determined by the dark states, i.e., in-phase in the presence of $|\psi_{D,2}\rangle$, anti-phase in the presence of $|\psi_{D,2}^{\prime\prime}\rangle$.

\begin{figure}
    \includegraphics[width=1\linewidth]{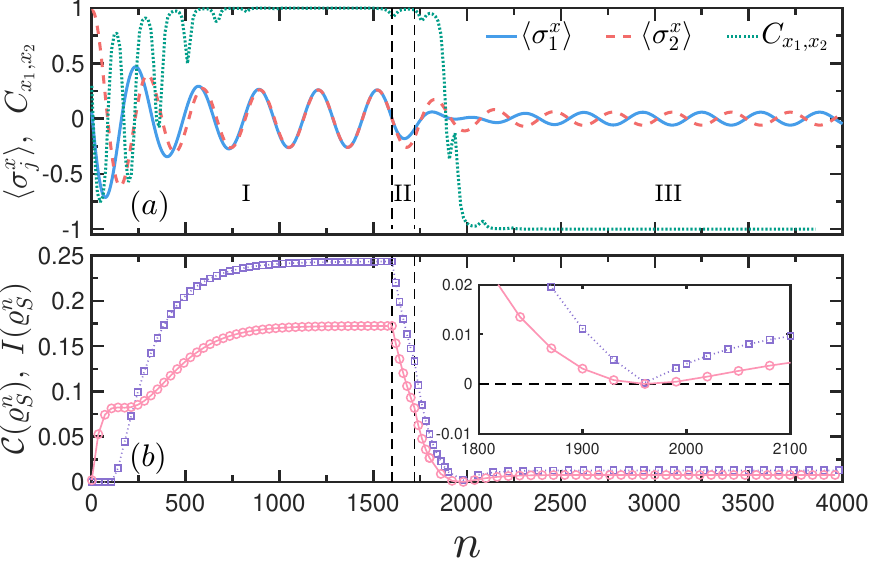}
    \caption{\label{Fig:CM} (a) The blue solid, red dashed, and green dotted lines show the order parameter $\langle\sigma^{x}_{1}\rangle$, $\langle\sigma^{x}_{2}\rangle$, and their Pearson correlation $C_{x_{1},x_{2}}$ as a function of the collision number $n$, respectively. As $n$ increases, ancilla state changes from $|\psi_{E}\rangle_{\text{I}}$ ({\it Phase} I) to $|\psi_{E}\rangle_{\text{II}}$ ({\it Phase} II) and $|\psi_{E}\rangle_{\text{III}}$  ({\it Phase} III).
(b) The circles and squares show mutual information $I(\varrho^{n}_{S})$ and entanglement $\Cs(\varrho^{n}_{S})$ as a function of $n$, respectively.
The inset shows the zoom-in where mutual information and entanglement reach zero.
Here $\omega\tau=0.01$, $g^{2}\tau=1$, sample length $\Delta n=140$, and the randomly selected initial state is $|\psi\rangle^{\text{ini}}_{S}=[0.8579 + 0.2631i,0.2774 + 0.3013i, 0.0222 - 0.1480i, 0.0428 - 0.0532i]^{\mathsf{T}}$ with $\mathsf{T}$ denotes the transpose. }    
\end{figure}

{\it Synchronization transitions.--} 
With the knowledge of correspondence between the targeting dissipator and the state of environment (ancilla), we now proceed to demonstrate the synchronization transition.
We initialize ``quantum metronomes" in a random separable state, and prepare the ancilla in \textit{Phase} I: $|\psi_{E}\rangle_{\text{I}}$, steering the ``metronomes'' towards achieving in-phase synchronization.

The solid and dashed lines in Fig.~\ref{Fig:CM}(a) show the evolution of order parameters $\langle\sigma^{x}_{1}\rangle$ and $\langle\sigma^{x}_{2}\rangle$, respectively. 
The system transits from initial asynchronous behavior ($n\lesssim700$) to synchronization, which is consistent with our theoretical prediction.
To transit to the anti-phase synchronization, the final ancilla state has to be $|\psi_{E}\rangle_{\text{III}}$.
However, the projection from the final state of the in-phase synchronization phase to the decoherence-free subspace of the anti-phase synchronization results in state $|\psi_{D,1}\rangle$ only. As we argued earlier, a single dark state cannot induce 
oscillation in the order parameters.
To circumvent this issue, we introduce an intermediate transition phase \textit{Phase} II, where the ancilla state is $|\psi_{E}\rangle_{\text{II}}=|g\rangle$.
The dark state of this phase has finite overlap with both $|\psi_{D,2}\rangle$ and $|\psi_{D,2}''\rangle$, thus making the transition from in-phase to anti-phase synchronization possible.
As shown in Fig.~\ref{Fig:CM}(a), we quench the ancillas into \textit{Phase} II after $n=1600$ with $|\psi_{E}\rangle_{\text{II}}=|g\rangle$, and further quench the ancilla states to \textit{Phase} III at $n=1720$ to drive the system from in-phase synchronization to anti-phase synchronization.

To quantitatively assess the synchronization transition, we calculate the Pearson correlation coefficient $C_{x_{1},x_{2}}$ between the two order parameters as a function of $n$,~\cite{KarpatPRA2019,KarpatPRA2020,Mahlow2024May}
\begin{equation}
C_{\alpha,\,\beta}(n)=\frac{\sum_{j=n}^{n+\Delta n }(\alpha_{j}-\bar{\alpha})(\beta_{j}-\bar{\beta})}{\sqrt{\sum_{j=n}^{n+\Delta n}(\alpha_{j}-\bar{\alpha})^{2}\sum_{j=n}^{n+\Delta n}(\beta_{j}-\bar{\beta})^{2}}},   
\end{equation}
i.e., the ratio of the covariance to the respective standard deviations of discrete-time series data; here $\bar{\alpha}=\Delta n^{-1}\sum_{j=n}^{n+\Delta n}\alpha_{j}$ and $\bar{\beta}=\Delta n^{-1}\sum_{j=n}^{n+\Delta n}\beta_{j}$ denote the temporal averages, where $\Delta n$ is the sample length of the discrete data.
The green dashed line in Fig.~\ref{Fig:CM}(a) illustrates the progression of the Pearson correlation from initial unstable values to $1$ (in-phase synchronization) and eventually to $-1$ (anti-phase synchronization).

The establishment of quantum synchronization typically involves entanglement generation.
To quantify entanglement, we adopt one of the popular choice for two-level systems, the concurrence~\cite{Wootters1998PRL} $\Cs(\varrho_{S}^{n})\equiv\max(0,\sqrt{\lambda_{1}}-\sqrt{\lambda_{2}}-\sqrt{\lambda_{3}}-\sqrt{\lambda_{4}})$, where $\lambda_{j}$ is the eigenvalue of $\varrho^{n}_{S}\tilde{\varrho}^{n}_{S}$ in decreasing order, and $\tilde{\varrho}^{n}_{S}$ is the spin-flipped state, defined as $\tilde{\varrho}^{n}_{S}=(\sigma_y \otimes \sigma_y) \varrho^{n*}_S (\sigma_y \otimes \sigma_y)$, where $\varrho^{n*}_{S}$ is the complex conjugate of $\varrho^{n}_{S}$ and $\sigma^y$ is the Pauli matrix.
Concurrence takes values in the range $[0,1]$, where $0$ indicates a separable state and $1$ represents a maximally entangled state.
Concurrence is defined for both pure and mixed states and is non-increasing under local operations and classical communication~\cite{Nielsen1999}, making it one of the ideal quantifiers of entanglement in two-qubit systems.
Meanwhile, correlations are assessed using the mutual information $I(\varrho^{n}_{S})\equiv S(\varrho^{n}_{S,1})+S(\varrho^{n}_{S,2})-S(\varrho^{n}_{S})$, where $S(\cdot)$ is the von Neumann entropy and $\varrho_{S,j}^{n}=\text{tr}_{k,k\neq j}[\varrho^{n}_{S}]$ is the reduced density matrix.

The squares and circles in Fig.~\ref{Fig:CM}(b) show the concurrence and mutual information, respectively, as a function of the number of collisions $n$. In {\it Phase} I, mutual information emerges rapidly and reaches a plateau, while entanglement generation lags. 
This is due to the fact that concurrence is a measure of quantum correlations, while mutual information encompasses both classical and quantum correlations. 
That is, classical correlations develop first. As entanglement increases, synchronization gradually establishes, which further boosts the rise of mutual information again until both reach a plateau.
In {\it Phase} II, the ancilla quenching decouples the ``metronomes", causing both measures to decay.
In {\it Phase} III, both correlations continue to decline, eventually vanish, and then revive.

We now explain the mechanism for the delayed vanishing and revival of the entanglement by a three-level system. Due to the absence of coherent pumping in Eq.~(\ref{Eq:ME}), the system can be simplified by adiabatically eliminating the rapidly decaying excited state $|\uparrow\uparrow\rangle$. 
The resulting Hilbert space is exactly the union of all the decoherence-free subspaces at different phases.
Within this subspace, the concurrence of the system is analytically tractable: $\Cs(\varrho_{S})=\left|c_0\text{tr}[|\downarrow\uparrow\rangle\langle\uparrow\downarrow|\varrho_{S}]\right|$, where $|\cdot|$ denotes absolute value and $c_0$ is the proportionality constant~\cite{Supple}.
The revival stems from sign changes in $c_0$ during in-phase/anti-phase transitions, as the population shifts from the symmetric to antisymmetric state, resulting in temporary nullification of entanglement.

\begin{figure}
    \includegraphics[width=1\linewidth]{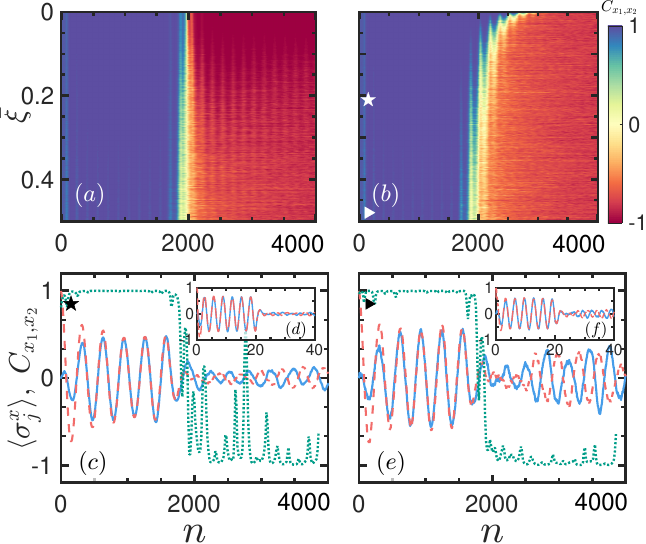}
    \caption{\label{Fig:Noise} Comparison of two ancilla quench sequences under noise: Panel (a) and (b) are for the complete sequence ({\it Phase} I $\to$II$\to$III) and the sequence without {\it Phase} II, respectively. 
    Both panels display Pearson correlation $C_{x_{1},x_{2}}$ as a function of collision number $n$ and noise strength $\bar{\xi}$, with sample length $\Delta n=140$. 
    The blue solid, red dashed, and green dotted lines in panels (b) and (e) show $\langle\sigma^{x}_{1}\rangle$, $\langle\sigma^{x}_{2}\rangle$, and $C_{x_{1},x_{2}}$ as a function of the collision number $n$ for $\bar{\xi}=0.21$ and $\bar{\xi}=0.49$, respectively, in the sequence without {\it Phase} II.
    The blue solid and red dashed lines in insets (d) and (f) show $\langle\sigma^{x}_{1}\rangle_{\text{traj}}$ and $\langle\sigma^{x}_{2}\rangle_{\text{traj}}$ along a single quantum trajectory, obtained from \SSE.
The other parameters are the same as Fig.~\ref{Fig:CM}.}  
\end{figure}

{\it Robustness to noise.--} 
To test the noise robustness, we introduce an additional stochastic term $H_{\text{noise}}=\sum_{j=1,2}\hbar\xi_{j}^{\alpha}(n)\sigma^{\alpha}_{j}$, where $\xi_{j}^{\alpha}(n)$ ($\alpha=x,y$) represents random fluctuations within $[-\bar{\xi},\bar{\xi}]$ at collision number $n$, where $\bar{\xi}$ controls the strength.

For controlled comparison across noise strength, we initialize the system to $|\psi\rangle_{S}^{\text{ini}}=\frac{|\uparrow\uparrow\rangle+|\downarrow\uparrow\rangle}{\sqrt{2}}$, purposely avoiding initial in-phase dynamics.
Interestingly, since the noise could induce spontaneous departure from the decoherence-free subspace, \textit{Phase} II is redundant for synchronization transitions.
To double check, we compare the complete sequence ({\it Phase} I$\to$II$\to$III) [Fig.~\ref{Fig:Noise}(a)] with the sequence without {\it Phase} II, i.e., ({\it Phase} I$\to$III) [Fig.~\ref{Fig:Noise} (b)].

Figures~\ref{Fig:Noise}(a)-(b) show the density plot of the Pearson correlation as a function of collision number $n$ and noise strength $\bar{\xi}$. 
The transition is always clear regardless of noise strength, except that small enough noise leads to a delay in the transition for the sequence without {\it Phase} II [see Fig.~\ref{Fig:Noise}(b)]. In the zero-noise limit, the transition occurs at $n=\infty$, which is consistent with the theoretical prediction.

We illustrate the evolution of order parameters and Pearson correlation for the sequence without {\it Phase} II under weak ($\bar{\xi}=0.21$) and strong ($\bar{\xi}=0.49$) noise in Figs.~\ref{Fig:Noise}(c) and (e), respectively. 
The enhanced Pearson correlation with strong noise indicates that noise do help with the anti-phase transition.
Notably, this robustness persists not only in ensemble averages but also at the single-trajectory level, as calculated via the \SSE~\cite{Supple,SchmolkePRL2024, SaniPRR2020,xingliPRA2023,liBohmPRA2022} [Figs.~\ref{Fig:Noise}(d) and (f)].
The elimination of {\it Phase} II is favored since it reduces the total time needed for a quantum simulation [See SM~\cite{Supple} for discussion of the complete sequence]. 
\begin{figure}
    \includegraphics[width=1\linewidth]{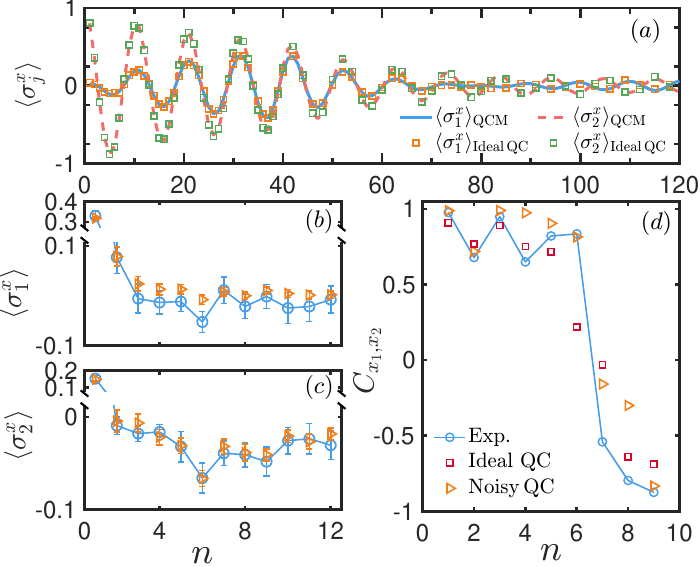}
       \caption{\label{Fig:Exper} (a) The lines and squares show $\langle \sigma^{x}_{j}\rangle$ as a function of collision number $n$ for the quantum collision model (denoted by QCM) and ideal quantum circuits calculation (denoted by Ideal QC), respectively, with $\omega\tau=0.3$ and $g^{2}\tau = 1$.
(b)-(c) The circles and triangles with error bar show $\langle\sigma^{x}_{j}\rangle$ as a function of $n$ computed via superconducting circuit simulation and noisy quantum circuits calculation, respectively, with $\omega\tau=0.8$ and $g^{2}\tau = 4$.
(d) The blue circles, red squares, and yellow triangles show the Pearson correlation $C_{x_{1},x_{2}}$ as a function of $n$ computed via superconducting circuit simulation (Exp.), ideal quantum circuits calculation (Ideal QC), and noisy quantum circuits calculation (Noisy QC), respectively, with a sample length $\Delta n =4$.
}

\end{figure}

{\it Experimental realization.--} 
We now turn to the experimental realization of our model on quantum circuits made of qubits. While the systems are made of two qubits, the ancilla states which consist of three energy levels need two qubits to encode.
Thus, we need to first map our system-ancilla interaction model onto a quantum circuit model by encoding the ancilla states as $|g\rangle\mapsto|\downarrow\downarrow\rangle_{\text{QC}}$, $|e\rangle\mapsto|\downarrow\uparrow\rangle_{\text{QC}}$, and $|r\rangle\mapsto|\uparrow\downarrow\rangle_{\text{QC}}$.
The ancilla states are reset to the initial state after each collision with the two spins. 
We show the quantum circuit for our QCM in Fig.~\ref{Fig:Sketch}(c).
Our QCM minimally requires four logical qubits: system qubits $s_{0}$, $s_{1}$ and ancilla qubits $a_{0}$, $a_{1}$, with a connectivity topology of $\tikz[baseline=(a.base),every node/.style={inner sep=0.2pt},x=1.8em]
{
  \node (a) at (0,0) {$s_0$};
  \node (b) at (1,0) {$a_0$};
  \node (c) at (2,0) {$a_1$};
  \node (d) at (3,0) {$s_1$};
  \draw (a) -- (b) -- (c) -- (d);
  \draw[bend left=18] (a.north) to (c.north);
  \draw[bend left=-18] (b.south) to (d.south);
}.$
We realize the simultaneous system-ancilla interactions via the second order Suzuki-Trotter decomposition~\cite{suzuki} (See SM~\cite{Supple} for detailed implementation).

The lines and squares in Fig.~\ref{Fig:Exper}(a) shows the order parameters $\langle\sigma^{x}_{j}\rangle$ as a function of $n$ for the quantum collision model and ideal quantum circuits~\cite{Qiskit} .
The excellent agreement validates the effectiveness of our mapping scheme.
However, the cumulative errors from hardware noise and decoherence increase exponentially with the circuit depth, which is proportional to collision number $n$.
To mitigate, we accelerate the system evolution such that $n\leq12$ by setting $\omega\tau=0.8$ and $g^{2}\tau=4$, utilize the quench sequence without intermediate {\it Phase} II and triggering ancilla quench at $n=5$, and initialize the system in a nearly in-phase synchronized state.

The circles with error bar in Figs.~\ref{Fig:Exper}(b)-(c) show the order parameters $\langle\sigma^{x}_{j}\rangle$ as a function of $n$ for the quantum simulation on the superconducting circuits~ (See detailed implementation in SM~\cite{Supple}.)
Due to the destructive nature of quantum measurements, for each $n$, the experimental sequence start from the initial state. 
To ensure statistical significance, for each data point, we repeat 25 times, each comprising 5000 shots of quantum circuit executions.
We observe that the oscillation amplitude decays rapidly at first but stabilizes within a narrow range. 
Though the noise significantly affects the oscillation amplitude, we still observe that a weak in-phase ($n<6$) and anti-phase ($n>6$) oscillation. This is clearly shown by the Pearson correlation $C_{x_{1},x_{2}}$ [circles in Fig.~\ref{Fig:Exper}(d)]. 
Our experimental results are largely consistent with noisy quantum circuit simulations (triangles in Fig.~\ref{Fig:Exper}), which mimic realistic conditions by applying various noise channels, such as depolarizing or relaxation errors, to each gate operation. We expect that future simulations on quantum computers with a more suitable topology, such as qutrit-based circuits~\cite{Morvan2021May, Goss2022Dec, Luo2023Jan, Roy2023Jun} could potentially yield improved results.

{\it Conclusions.--} 
In summary, we have established a quantum analog of classical synchronization transition by engineering the state of ancilla, i.e., the environment of two noninteracting qubits.
We have shown that the two-body dissipator can be engineered by quenching ancilla states, which leads to a transition from in-phase to anti-phase synchronization.
This transition is robust against noise, and we have observed the signatures of synchronization transition from the Pearson correlation in quantum circuits simulations.
Note that here, since the qubits does not exhibits nonlinearity, there exists no limit cycle. We expect the synchronization transition to be more stable for nonlinear systems with limit cycles, such as Van der Pol oscillators~\cite{Lee2014Feb}.

Meanwhile, it is interesting to note that, by engineering the state of the ancilla, we obtained an effective master equation with a two-body dissipator with desired phase factors. 
Starting from a product state, we can eventually obtain a two-qubit entangled state by the dissipative coupling.
We have also observed the vanishing and reviving of entanglement and established the connection between entanglement behavior and synchronization transition. This is a step towards many-body dissipator engineering~\cite{Harrington2022Oct}, which is promising in realizing full control of many-body entangled states.

\begin{acknowledgments}
{\it Acknowledgments.--} 
This work is supported by the Quantum Science and Technology-National Science and Technology Major Project (2024ZD0300600).
We acknowledge financial support from 
the National Natural Science Foundation of China under Grant No. 92565105 and 12204395,
Hong Kong RGC No. 14301425, No. 24308323, and No. C4050-23GF,
Guangdong Provincial Quantum Science Strategic Initiative No. GDZX2404004 and GDZX2505005,
the Space Application System of China Manned Space Program, 
and CUHK Direct Grant.

\end{acknowledgments}

\bibliography{Ref}

\end{document}


\beginsupplement
\title{Supplementary Material for ``Signatures of Environment-Induced Quantum Synchronization Transitions via Two-body Dissipator Engineering"}

\author{Xingli Li}
\affiliation{Department of Physics, The Chinese University of Hong Kong, Shatin, New Territories, Hong Kong, China}

\author{Yan Li}
\affiliation{Department of Physics, The Chinese University of Hong Kong, Shatin, New Territories, Hong Kong, China}

\author{Yangqian Yan}
\email{yqyan@cuhk.edu.hk}
\affiliation{Department of Physics, The Chinese University of Hong Kong, Shatin, New Territories, Hong Kong, China}
\affiliation{State Key Laboratory of Quantum Information Technologies and Materials, The Chinese University of Hong Kong, Hong Kong SAR, China}
\affiliation{The Chinese University of Hong Kong Shenzhen Research Institute, 518057 Shenzhen, China}

\maketitle
\tableofcontents

\section{The derivation of quantum master equation in quantum collision model framework}
\label{SezcI}

Quantum collision model is inspired by Boltzmann's {\it Stosszahlansatz}~\cite{BrownSHPSB2009}, which conceptualizes an open quantum system as the system $S$ ``collides'' with only a fraction of the degrees of freedom of the environment over a small duration $\tau$~\cite{CiccarelloPR2022}.
Building upon this ansatz, the environment is discretized into a sequence of individual components, denoted by states $\eta^{n}_{E}$, ($n=1,2,\dots$). We incorporate the stroboscopic interaction between the system and environmental by a unitary operator $U=e^{-iH\tau/\hbar}$ [Fig.~1(b) in the main text].
After the $n$-th collision, the system state $\varrho^{n}_{S}$ evolves into $\varrho^{n+1}_{S}$, 
\begin{equation}
\varrho^{n}_{S}\mapsto\varrho^{n+1}_{S}=\text{tr}_{\eta^{n+1}_{E}}[U\varrho^{n}_{S}\otimes\eta^{n+1}_{E}U^{\dagger}]\coloneqq\Phi[\varrho^{n}_{S}],
\label{Eq:CPTPmap}
\end{equation}
where $\Phi[\cdot]$ is a completely positive trace-preserving map.
In continuous-time limit ($\tau\to0$), Taylor expanding Eq.~(\ref{Eq:CPTPmap}) leads to a Lindblad master equation
\begin{equation}
\lim_{\tau\to0}\frac{\varrho^{n+1}_{S}-\varrho^{n}_{S}}{\tau} = \frac{d\varrho_{S}}{dt}=-\frac{i}{\hbar}[H,\varrho_{S}]+\gamma\Ds_{\varrho_{S}}[o],
\label{Eq:OME}
\end{equation}
where $\Ds_{\varrho_{S}}[o] = o\varrho_{S}o^{\dagger}-\frac{1}{2}(o^{\dagger}o\varrho_{S}+\varrho_{S}o^{\dagger}o)$ is dissipator.
The decay rate $\gamma$ is determined by environment correlation function~\cite{BreuerBook2002}.

Based on the above theoretical framework, we present the detailed derivation of our effective master equation.
In our model, the unitary operator $U$ is composed of two parts, i.e., $U=U_{I}U_{S}$, where $U_{S}$ and $U_{I}$ represent the free evolution of the system and the interaction between the system and environment, respectively.
Therefore, we start the derivation with the corresponding unitary evolution operators.
The self-Hamiltonian of the non-interacting system reads
\begin{equation}
H_{S}=\hbar\omega(\sigma^{z}_{1}+\sigma^{z}_{2}),
\end{equation}
and the Hamiltonian that describes the interaction between the system block and one environment block (single  collision process) reads
\begin{equation}
\begin{aligned}
H_{I}=& g\hbar\sum_{j}V_{S,j}\otimes V_{E,j}^{\dagger}+\text{H.c.}, \\
\end{aligned}
\end{equation}
where $V_{S,j}$ and $V_{E,j}$ represent the system's operator and environment's operator in the $j$-th component of the interaction Hamiltonian with the interaction strength $g$, respectively. Based on the above, the corresponding unitary time-evolution operators reads $U_{S}=e^{-i\hat{H}_{S}\tau_{S}/\hbar}$ and $U_{I}=e^{-i\hat{H}_{I}\tau_{I}/\hbar}$. 
For simplicity, We assume that the time interval for the system's free evolution is the same as the time interval for the interaction between the system and the environment, i.e., $\tau_{S}=\tau_{I}=\tau$.

\begin{figure}[!htpb]
    \includegraphics[width=1\textwidth]{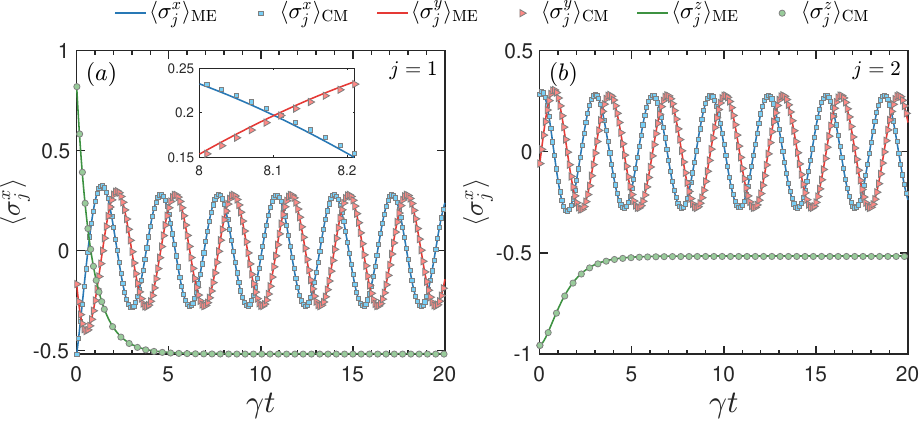}
    \caption{\label{Fig:Comparison} Comparison of the time-evolution expectation values of local observables $\sigma^{\alpha}_{j}$, ($\alpha=x,y,z$) for two non-interacting spins (a) $j=1$, (b) $j=2$ based on master equation (solid lines) and quantum collision model (markers). The zoom-in of (a) indicates the partial details of time evolution with $\gamma t\in[8,8.2]$. The other parameters are set as follows: $\omega\tau=0.02$ and $g=10$ for collision model and  $\gamma=1$ for Lindblad master equation.}    
\end{figure} 

In the continuous-time limit ($\tau\to0$), we expand the unitary time-evolution operators into the Taylor series as follows:
\begin{equation}
\begin{aligned}
U_{S}=&\mathbb{I}-\frac{i\tau\omega}{\hbar}(\sigma^{z}_{1}+\sigma^{z}_{2})-\frac{\tau^{2}\omega^2}{2\hbar^2}(\sigma^{z}_{1}+\sigma^{z}_{2})^2+o(\tau^{n}),\\
U_{I}=&\mathbb{I}-\frac{ig\tau}{\hbar}\sum_{j}[V_{S,j}\otimes V^{\dagger}_{E,j}+\text{H.c.}]-\frac{g^{2}\tau^{2}}{2\hbar^2}\sum_{j,k}[(V_{S,j}\otimes V^{\dagger}_{E,j}+\text{H. c.})(V_{S,k}\otimes V^{\dagger}_{E,k}+\text{H.c.})]+o(\tau^{n}),
\end{aligned}    
\end{equation}
where $o(\tau)$ indicates the high-order terms. 
Further assuming that 
(i) the environment blocks do not interact with each other, 
(ii) the environment blocks are initially uncorrelated, and 
(iii) each environment block collides with the system block only once~\cite{BreuerBook2002}, we utilize the Markov approximation to derive the mapping of the processes from the $n$-th collision to the $n+1$-th collision as follows
\begin{equation}
\begin{aligned}
\varrho^{n}_{S}\mapsto \varrho^{n+1}_{S}=&\text{tr}_{\eta^{n+1}_{E}}[U_{I}U_{S}(\varrho^{n}_{S}\otimes\eta^{n+1}_{E})U_{S}^{\dagger}U_{I}^{\dagger}],\\   
 =&\text{tr}_{\eta^{n+1}_{E}}[(\mathbb{I}-\frac{ig\tau}{\hbar} H_{I}-\frac{g^{2}\tau^{2}}{2\hbar^2}H_{I}^{2})(\mathbb{I}-\frac{i\tau}{\hbar} H_{S}-\frac{\tau^{2}}{2\hbar^2}H_{S}^{2})(\varrho^{n}_{S}\otimes\eta^{n+1}_{E})(\mathbb{I}+\frac{i\tau}{\hbar} H_{S}-\frac{\tau^{2}}{2\hbar^2} H_{S}^{2})(\mathbb{I}+\frac{ig\tau}{\hbar} H_{I}-\frac{g^{2}\tau^{2}}{2\hbar^2} H_{I}^{2})],\\
 =&\text{tr}_{\eta^{n+1}_{E}}[(\mathbb{I}-\frac{ig\tau}{\hbar} H_{I}-\frac{i\tau}{\hbar} H_{S}-\frac{g\tau^{2}}{\hbar^2}H_{I}H_{S}-\frac{g^{2}\tau^{2}}{2\hbar^2}H_{I}^{2})(\varrho^{n}_{S}\otimes\eta^{n+1}_{E})(\mathbb{I}+\frac{ig\tau}{\hbar}H_{I}+\frac{i\tau}{\hbar} H_{S}-\frac{g\tau^{2}}{\hbar^2}H_{S}H_{I}-\frac{g^{2}\tau^{2}}{2\hbar^2}H_{I}^{2})+g^{m}\cdot o(\tau^{n})],
\end{aligned}    
\end{equation}
where $\varrho^{n}_{S}$ is the density matrix of the system after $n$ times collision, and $\eta^{n}_{E}$ is the $n$-th state of the environment.

Recalling the timescale $\tau\to 0$ and assuming that $g^2\tau$ remains constant, we neglect terms of higher order terms. 
Specifically, we neglect terms of the form $g^{m}\cdot o(\tau^{n})$ where $n>m/2$ and assume the environment states satisfies $\text{tr}[\eta^{n}_{E}H_{I}]=0$. 
After this, we arrive
\begin{equation}
\begin{aligned}
\varrho^{n+1}_{S}=&\varrho^{n}_{S}-\frac{i\tau}{\hbar}\text{tr}_{\eta^{n+1}_{E}}[H_{S}(\varrho^{n}_{S}\otimes\eta^{n+1}_{E})-(\varrho^{n}_{S}\otimes\eta^{n+1}_{E})H_{S}]+\frac{g^2\tau^2}{\hbar^2}\text{tr}_{\eta^{n+1}_{E}}[H_{I}(\varrho^{n}_{s}\otimes\eta^{n+1}_{E})H_{I}-H_{I}^2(\varrho^{n}_{S}\otimes\eta^{n+1}_{E})/2-(\varrho^{n}_{S}\otimes\eta^{n+1}_{E})H_{I}^2/2],\\
=&\varrho^{n}_{S}-\frac{i\tau}{\hbar}(H_{S}\varrho^{n}_{S}-\varrho^{n}_{S}H_{S})+\frac{g^2\tau^2}{\hbar^2}\left\{\sum_{j,k}\text{tr}[V^{\dagger}_{E,j}V^{\dagger}_{E,k}\eta^{n+1}_{E}](V_{S,k}\varrho^{n}_{S}V_{S,j}-V_{S,j}V_{S,k}\varrho^{n}_{S}/2-\varrho^{n}_{S}V_{S,j}V_{S,k}/2) \right. \\ 
&\left.
+\sum_{j,k}\text{tr}[V_{E,j}V^{\dagger}_{E,k}\eta^{n+1}_{E}](V_{S,k}\varrho^{n}_{S}V_{S,j}^{\dagger}-V_{S,j}^{\dagger}V_{S,k}\varrho^{n}_{S}/2-\varrho^{n}_{S}V_{S,j}^{\dagger}V_{S,k}/2)\right.\\ 
&\left.
+\sum_{j,k}\text{tr}[V^{\dagger}_{E,j}V_{E,k}\eta^{n+1}_{E}](V_{S,k}^{\dagger}\varrho^{n}_{S}V_{S,j}-V_{S,j}V_{S,k}^{\dagger}\varrho^{n}_{S}/2-\varrho^{n}_{S}V_{S,j}V_{S,k}^{\dagger}/2)\right. \\ \
&\left.
+\sum_{j,k}\text{tr}[V_{E,j}V_{E,k}\eta^{n+1}_{E}](V_{S,k}^{\dagger}\varrho^{n}_{S}V_{S,j}^{\dagger}-V_{S,j}^{\dagger}V_{S,k}^{\dagger}\varrho^{n}_{S}/2-\varrho^{n}_{S}V_{S,j}^{\dagger}V_{S,k}^{\dagger}/2)
\right\}.
\end{aligned}
\label{Eq:DetailMap}
\end{equation}
In this work, the environment state is modeled by a system with three energy levels, namely $\{|g\rangle,|e\rangle,|r\rangle\}$. 
Then the environment operators $V_{E,j}$ can be determined specifically as $V_{E,1}=|g\rangle \langle r|$ and $V_{E,2}=|e\rangle \langle r|$. 
The system operators in interaction Hamiltonian, denoted as $V_{S,j}$, are $V_{S,1}=\sigma^{-}_{1}$ and $V_{S,2}=\sigma^{-}_{2}$, respectively.

Combing with Eq.~(\ref{Eq:DetailMap}), we obtain the Lindblad-form master equation 
\begin{equation}
\begin{aligned}
\frac{d\varrho_{S}}{dt}=&\lim_{\tau\to0}\frac{\varrho^{n+1}_{S}-\varrho^{n}_{S}}{\tau}\\
=&-\frac{i}{\hbar}[H_{S},\varrho_{S}]+\frac{g^{2}\tau}{\hbar^2}
\left\{\eta^{n+1}_{E|g,g}\cdot(\sigma^{-}_{1}\varrho_{S}\sigma^{+}_{1}-\sigma^{+}_{1}\sigma^{-}_{1}\varrho_{S}/2-\varrho_{S}\sigma^{+}_{1}\sigma^{-}_{1}/2) \right. \\
&\left.
+\eta^{n+1}_{E|g,e}\cdot(\sigma^{-}_{2}\varrho^{n}_{S}\sigma^{+}_{1}-\sigma^{+}_{1}\sigma^{-}_{2}\varrho^{n}_{S}/2-\varrho^{n}_{S}\sigma^{+}_{1}\sigma^{-}_{2}/2) \right.\\
&\left.
+\eta^{n+1}_{E|e,g}\cdot(\sigma^{-}_{1}\varrho_{S}\sigma^{+}_{2}-\sigma^{+}_{2}\sigma^{-}_{1}\varrho_{S}/2-\varrho_{S}\sigma^{+}_{2}\sigma^{-}_{1}/2)\right. \\
&\left.
+\eta^{n+1}_{E|e,e}\cdot(\sigma^{-}_{2}\varrho_{S}\sigma^{+}_{2}-\sigma^{+}_{2}\sigma^{-}_{2}\varrho_{S}/2-\varrho_{S}\sigma^{+}_{2}\sigma^{-}_{2}/2)\right. \\
&\left.
+\eta^{n+1}_{E|r,r}\cdot(\sigma^{+}_{1}\varrho_{S}\sigma^{-}_{1}-\sigma^{-}_{1}\sigma^{+}_{1}\varrho_{S}/2-\varrho_{S}\sigma^{-}_{1}\sigma^{+}_{1}/2) \right.\\
&\left.
+\eta^{n+1}_{E|r,r}\cdot(\sigma^{+}_{2}\varrho_{S}\sigma^{-}_{2}-\sigma^{-}_{2}\sigma^{+}_{2}\varrho_{S}/2-\varrho_{S}\sigma^{-}_{2}\sigma^{+}_{2}/2)
\right\},   
\end{aligned}   
\label{Eq:CMtoME}
\end{equation}
where $\eta^{n+1}_{E|\alpha,\beta}$ is the density matrix element of the ancilla state ($\alpha,\beta\in[g,e,r]$). The coefficient $\eta^{n+1}_{E|\alpha,\beta}\cdot g^{2}\tau$ is the decay rate $\gamma_{\alpha,\beta}$ for each loss or gain channel. 
Recalling the ancilla state given in the main text, i.e., $\eta_{E}=|\psi_{E}\rangle\langle\psi_{E}|$, where $|\psi_{E}\rangle=\cos{\theta}|g\rangle+\sin{\theta}e^{i\phi}|e\rangle$, Eq.~(\ref{Eq:CMtoME}) can be simplified as
\begin{equation}
\begin{aligned}
\frac{d\varrho_{S}}{dt}=&-\frac{i}{\hbar}[H_{S},\varrho_{S}]+\frac{g^{2}\tau}{\hbar^2}
\left\{\cos^2{\theta}\cdot(\sigma^{-}_{1}\varrho_{S}\sigma^{+}_{1}-\sigma^{+}_{1}\sigma^{-}_{1}\varrho_{S}/2-\varrho_{S}\sigma^{+}_{1}\sigma^{-}_{1}/2) \right. \\
&\left.
+\cos{\theta}\sin{\theta}e^{i\phi}\cdot(\sigma^{-}_{2}\varrho_{S}\sigma^{+}_{1}-\sigma^{+}_{1}\sigma^{-}_{2}\varrho_{S}/2-\varrho_{S}\sigma^{+}_{1}\sigma^{-}_{2}/2) \right.\\
&\left.
+\cos{\theta}\sin{\theta}e^{-i\phi}\cdot(\sigma^{-}_{1}\varrho_{S}\sigma^{+}_{2}-\sigma^{+}_{2}\sigma^{-}_{1}\varrho_{S}/2-\varrho_{S}\sigma^{+}_{2}\sigma^{-}_{1}/2)\right. \\
&\left.
+\sin^2{\theta}\cdot(\sigma^{-}_{2}\varrho_{S}\sigma^{+}_{2}-\sigma^{+}_{2}\sigma^{-}_{2}\varrho_{S}/2-\varrho_{S}\sigma^{+}_{2}\sigma^{-}_{2}/2)
\right\},\\
=&-\frac{i}{\hbar}[H_{S},\varrho_{S}]+\frac{g^{2}\tau}{\hbar^2}\Ds_{\varrho_{S}}[\cos{\theta}\sigma^{-}_{1}+\sin{\theta}e^{i\phi}\sigma^{-}_{2}].\\
\end{aligned}
\label{S1eq:me}
\end{equation}
Thus, we obtain the effective master equation shown in Eq.~(1) in the main text.
To double check, we perform a simulation for ancilla state $\eta^{n}_{E}=\eta^{n+1}_{E}=(|g\rangle\langle g|+|e\rangle\langle e| + |g\rangle\langle e| + |e\rangle\langle g|)/2$. 
The effective non-zero decay rates reads $\gamma_{g,g}=\gamma_{e,e}=\gamma_{e,g}=\gamma_{g,e}=1$ for $g^{2}\tau=2$. 
The jump operators reads $o=\sigma^{-}_{1} + \sigma^{-}_{2}$. The solid lines and symbols in Fig.~\ref{Fig:Comparison} show the time-evolution expectation values of local observables $\sigma^{\alpha}_{j}$, ($\alpha=x,y,z$) for two non-interacting spins based on master equation and quantum collision model, respectively.

\section{Alternative theoretical framework: ancilla with large spontaneous emission}

In the quantum collision model, the environment is modeled by a sequence of ancillary systems that interact with the system sequentially.
Here we offer an alternative framework to derive the effective quantum master equation by considering the ancilla with large spontaneous emission.
We still consider two qubits interact with a three-level ancilla, characterized by $H=H_{S}+H_{I}$.
The ancilla is kept unchanged instead of updating it via Eq.~(\ref{Eq:CPTPmap}), but experiences simultaneously intense spontaneous emissions from $|r\rangle\to|e\rangle$ and $|r\rangle\to|g\rangle$ with identical emissivity,  resulting in the following master equation:
\begin{equation}
  \frac{d\varrho}{dt}=-\frac{i}{\hbar}[H,\varrho]+\gamma\Ds_{\varrho}[|g\rangle\langle r|+|e\rangle\langle r|],
\label{Eq:APPOME}
\end{equation}
where $\gamma$ is the spontaneous emission rate.
Considering the decaying excited states $|r\rangle$ is almost unpopulated, we adiabatically eliminate the excited state under the assumption of a perturbative interaction.
Following Ref.~\cite{EffectiveOQSReiter2012Mar}, we decompose the interaction Hamiltonian as $H_{I}=H_{-}+H_{+}$, with $H_{+}=(H_{-})^{\dagger}=\hbar g(\sigma_{1}^{-}\otimes |r\rangle\langle g|+\sigma_{2}^{-}\otimes |r\rangle\langle e|)$.
Then we obtain the non-Hermitian Hamiltonian from Eq.~(\ref{Eq:APPOME}), which is $H_{\text{NH}}=-\frac{i\gamma}{2}o^{\dagger}o=-i\gamma|r\rangle\langle r|$.
Omitting the self-energy of the ancilla, we obtain an effective 
master equation, $ \frac{d \varrho}{dt}=-\frac{i}{\hbar}[H_{\text{eff}},\varrho]+\Ds_{\varrho}[o_{\text{eff}}]$, where the effective Hamiltonian is $H_{\text{eff}}=H_{S}-\frac{1}{2}[H_{-}(H_{\text{NH}})^{-1}H_{+} + \text{H.c.}]=H_{S}$ and the effective jump operator is
\begin{equation}
\begin{aligned}
o_{\text{eff}}=&\sqrt{\gamma}(|g\rangle\langle r|+|e\rangle\langle r|)(H_{\text{NH}})^{-1}H_{+},\\
=&\frac{ig}{\sqrt{\gamma}}[\sigma^{-}_{1}\otimes(|e\rangle\langle g|+|g\rangle\langle g|)+\sigma^{-}_{2}\otimes(|e\rangle\langle e|+|g\rangle\langle e|)].
\label{Eq:APPo}      
\end{aligned}
\end{equation}
Substituting the ancilla state $|\psi_{E}\rangle=\cos{\theta}|g\rangle+\sin{\theta}e^{i\phi}|e\rangle$ into Eq.~(\ref{Eq:APPo}), we obtain $o_{\text{eff}}=\frac{ig}{\sqrt{2\gamma}}(\cos{\theta}\sigma^{-}_{1}+\sin{\theta}e^{i\phi}\sigma^{-}_{2})\otimes(|e\rangle + |g\rangle)$. Tracing out the ancilla state, we then reproduces Eq.~(\ref{S1eq:me}).

\section{Elimination of the excited state}

In the main text, we stated that since there is an absence of coherent pumping between the energy levels of the system, the presence of dissipation will lead to the population of excited state decaying rapidly to zero. 
Therefore, it is reasonable to directly eliminate the excited state and reduce the system to a three-level system. 
Here we present the numerical results obtained by the full and reduced master equations to verify this point.

The full master equation is given by Eq.~(\ref{S1eq:me}), and the reduced master equation is described in the following way
\begin{equation}
\frac{d}{dt}\varrho_{S}^{D}=-\frac{i}{\hbar}[H_{S}^{D},\varrho_{S}]+\frac{g^{2}\tau}{\hbar^2}\Ds_{\varrho_{S}^{D}}[\frac{\cos{\theta}}{2}(\sigma^{-}_{1}-\sigma^{-}_{1}\sigma^{z}_{2})+\frac{\sin{\theta}e^{i\phi}}{2}(\sigma^{-}_{2}-\sigma^{z}_{1}\sigma^{-}_{2})],
\label{S2eq:rme}
\end{equation}
where the reduced self-Hamiltonian $H_{S}^{D}$ differs from $H_{S}$ in setting the excited-state self energy to zero. 
In addition, compared to Eq.~(\ref{S1eq:me}), the changes in the jump operator in Eq.~(\ref{S2eq:rme}) are also due to the need to eliminate all terms containing $|\uparrow\uparrow\rangle$ state. 
As a result, all operators in Eq.~(\ref{S2eq:rme}) are now restricted to the subspace expanded by states $\{|\downarrow\uparrow\rangle,|\uparrow \downarrow\rangle, |\downarrow\downarrow\rangle\}$.

As for the initial state $\varrho_{S}^{D}$ of the reduced master equation, we similarly eliminate all terms containing the $|\uparrow\uparrow\rangle$ state. 
Therefore, after giving an initial state $\varrho_{S}$ for the full master equation, we obtain $\varrho_{S}^{D}$ by setting first row and first column elements of $\varrho_{S}$ to zero, and combining the populations of $|\uparrow\uparrow\rangle$ and $|\downarrow\downarrow\rangle$ to maintain trace-preserving. Furthermore, it is imperative to confirm that $\varrho_{S}^{D}$ adheres to the requirement of semi-positive definiteness.

 \begin{figure}[!htpb]
    \centering
    \includegraphics[width=1\textwidth]{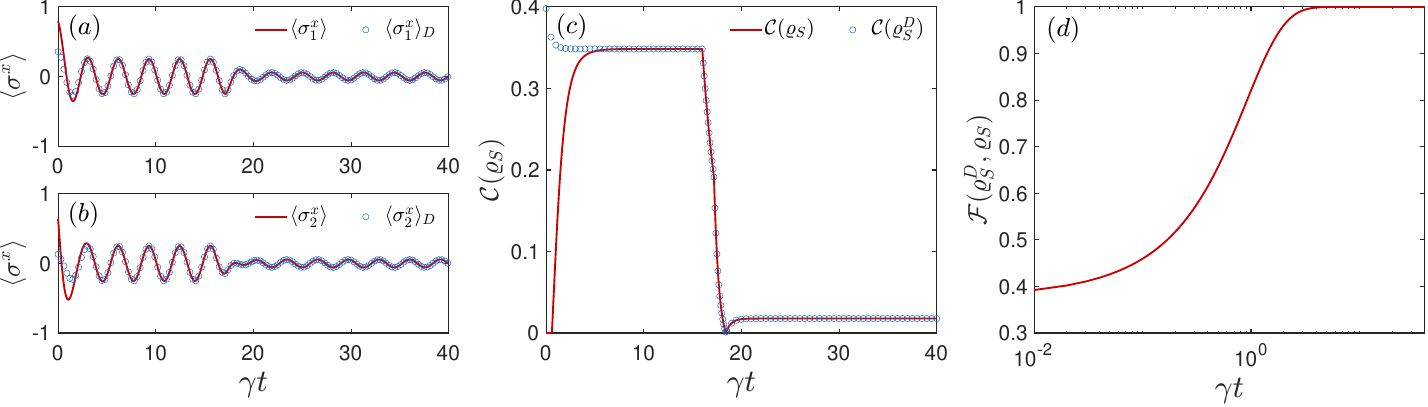} 
    \caption{\label{Fig:Eff} (a)-(b) The red solid lines and blue circles show order parameters $\langle\sigma^{x}_{j}\rangle$ and $\langle\sigma^{x}_{j}\rangle_{D}$ as a function of time $t$,  which are obtained by full and reduced master equation, respectively.  (c) Concurrence of $\varrho_{S}$ (red solid line) and $\varrho_{S}^{D}$ (blue circles) varies with time. (d) The red solid line shows fidelity between $\varrho_{S}$ and $\varrho_{S}^{D}$ changes over time. The other parameters are $\omega=1$, $\gamma=1$.}    
\end{figure}

In Figs.~\ref{Fig:Eff}(a) and (b), we show the order parameters $\langle\sigma^{x}_{j}\rangle$ (red solid lines) and $\langle\sigma^{x}_{j}\rangle_{D}$ (blue circles) as a function of time, which are obtained by full and reduced master equations, respectively. 
We can notice that there is a difference at the initial time, and as the evolution proceeds, the results gradually coincide. 
Besides, in Fig.~\ref{Fig:Eff}(c), the solid line and markers show the concurrence of $\varrho_{S}$ and $\varrho_{S}^{D}$ varies with the time, respectively. 
The coincidence between the solid line and markers in the long-time evolution shows that it is feasible to eliminate the $|\uparrow\uparrow\rangle$. 
The reduced master equation is valid in discussing the  quantities we are interested in.

Furthermore, we use a notion of quantum information theory, the well-known fidelity to measure the "closeness" of the two states. The fidelity between density matrices $\varrho_{1}$ and $\varrho_{2}$ is defined as follows~\cite{Fidelity}:
\begin{equation}
\Fs(\varrho_{1},\varrho_{2})= \left(\text{tr}
\left[\sqrt{\sqrt{\varrho_{1}}\varrho_{2}\sqrt{\varrho_{1}}}\right]\right)^2.  
\end{equation}
In Fig.~\ref{Fig:Eff}(d), the solid line shows fidelity $\Fs(\varrho_{S}^{D},\varrho_{S})$ as a function of time, which also suggest that the reduced master equation can effectively describe the properties of the original full master equation.

Since the reduced master equation is valid in discussing of entanglement (concurrence), we now analyze vanishing and revival of entanglement by $\varrho_{S}^{D}$.
The reduced density matrix $\varrho_{S}^{D}$ can be expressed in general form,
\begin{equation}
\varrho_{S}^{D}=
\begin{pmatrix}
0 & 0 &0 &0 \\
0 & \varrho_{22} & \varrho_{23} &\varrho_{24} \\ 
0 & \varrho_{32} & \varrho_{33} &\varrho_{34} \\ 
0 & \varrho_{42} & \varrho_{43} &\varrho_{44} 
\end{pmatrix}.    
\end{equation}
Following the definition of concurrence given in the main text, $\Cs(\varrho_{S}^{D})\equiv\max(0,\sqrt{\lambda_{1}}-\sqrt{\lambda_{2}}-\sqrt{\lambda_{3}}-\sqrt{\lambda_{4}})$,  where $\lambda_{1}\geq\lambda_{2}\geq\lambda_{3}\geq\lambda_{4}$ are the eigenvalue of $\varrho^{D}_{S}\tilde{\varrho}^{D}_{S}$, $\tilde{\varrho}^{D}_{S}=(\sigma_{y}\otimes\sigma_{y})(\varrho^{D}_{S})^{*}(\sigma_{y}\otimes\sigma_{y})$ and $*$ denotes complex conjugate operation. 
Then, we obtain $\sqrt{\lambda_{1}}=|\sqrt{\varrho_{22}\varrho_{33}}+\sqrt{\varrho_{23}\varrho_{32}}|$, $\sqrt{\lambda_{2}}=|\sqrt{\varrho_{22}\varrho_{33}}-\sqrt{\varrho_{23}\varrho_{32}}|$, $\sqrt{\lambda_{3}}=\sqrt{\lambda_{4}}=0$. 
Since the decoherence-free subspaces in the main text always contain the basis $|\downarrow\uparrow\rangle$, concurrence $\Cs(\varrho^{D}_{S})$ will reach zero when 
absence the information of the basis $|\uparrow\downarrow\rangle$ in $\varrho^{D}_{S}$, e.g., $\varrho_{22}=\varrho_{23}=0$. 
Therefore, the vanishing and revival of entanglement can be understood as follows:
Since there is a change from a symmetric dark state ($|\psi_{D,2}\rangle$) to an antisymmetric dark state ($|\psi_{D,2}\rangle''$) during the evolution, $\text{tr}[|\downarrow\uparrow\rangle\langle\uparrow\downarrow|\varrho_{S}]$ will continuously change from a positive to a negative value. 
When $\text{tr}[|\downarrow\uparrow\rangle\langle\uparrow\downarrow|\varrho_{S}]=0$, we have $\sqrt{\lambda_{1}}=\sqrt{\lambda_{2}}$ and concurrence will vanish. Therefore, the vanishing and revival of entanglement is due to the change of dark states and is related to synchronization properties.

\section{Dynamics under stochastic Schr\"{o}ndinger equation}

The previous discussions highlight the equivalence between our collision model and Lindblad master equation under certain assumptions and approximations. 
However, the dissipative dynamics of a quantum system can be viewed as a result of continuous measurements of a closed quantum system in some ways. 
Therefore, we can unveil the dissipative dynamics of the system in the trajectory level by the diffusive stochastic Schr\"{o}ndinger equation. 
The SSE can be obtained according to the homodyne detection and described in the following form~\cite{SSEbook,Wiseman2009Nov}:
\begin{equation}
    d|\psi\rangle= \left[-\frac{i}{\hbar}H_{\text{eff}}dt+ \gamma\sum_{k}\frac{\langle X_{k}\rangle_{\psi}}{2}(o_{k}-\frac{\langle X_{k}\rangle_{\psi}}{4})dt+\sqrt{\gamma}\sum_{k} (o_{k}-\frac{\langle X_{k}\rangle_{\psi}}{2})dW_{k}\right]|\psi\rangle,
\end{equation}
where $H_{\text{eff}}=H_{S}-i\gamma\sum_{k}o^{\dagger}_{k}o_{k}/2$ is the non-Hermitian Hamiltonian, the expectation values $\langle X_{k}\rangle_{\psi}=\langle \psi|X_{k}|\psi\rangle$ with $X_{k}=o^{\dagger}_{k}+o_{k}$ and $dW_{k}$ are stochastic Wiener increments describing the white noise contribution, which satisfying $dW_{k}dW_{l}=\delta_{k,l}dt$ and $\mathbb{E}(dW_{k})=0$.

To showcase the dissipative dynamic of the system, we specify the jump operator, for instance, $o= \sigma^{-}_{1}+\sigma^{-}_{2}$ accompanied by the decay rate $\gamma$. 
By representing initial wave function as $|\psi\rangle=[\phi_{1},\phi_{2},\phi_{3},\phi_{4}]^{\rm T}$, where $\phi_{j}\in \mathbb{C}$, we obtain the expectation value $\langle X\rangle_{\psi}=2\text{Re}[(\phi^{*}_{1}+\phi^{*}_{4})(\phi_{2}+\phi_{3})]$ and have the following expression
\begin{equation}
 \begin{aligned}
   \Delta\phi_{1}=& \left[(-\frac{i}{\hbar}H^{1,1}_{\text{eff}}-\frac{(\langle X\rangle_{\psi}^{2}+8)\gamma}{8})\Delta t-\frac{\sqrt{\gamma}\langle X\rangle_{\psi}}{2}\Delta W\right]\phi_{1},\\
   \Delta\phi_{2}=& (\frac{\gamma\langle X\rangle_{\psi}}{2}\Delta t+  \sqrt{\gamma}\Delta W) \phi_{1} -\left[(\frac{i}{\hbar}H^{2,2}_{\text{eff}}+\frac{(\langle X\rangle_{\psi}^{2}+4)\gamma}{8}))\Delta t+\frac{\sqrt{\gamma}\langle X\rangle_{\psi}}{2}\Delta W\right]\phi_{2} -\frac{\gamma}{2}\Delta t\phi_{3},\\
   \Delta\phi_{3}=& (\frac{\gamma\langle X\rangle_{\psi}}{2}\Delta t+  \sqrt{\gamma}\Delta W) \phi_{1} -\frac{\gamma}{2}\Delta t\phi_{2} -\left[(\frac{i}{\hbar}H^{3,3}_{\text{eff}}+\frac{(\langle X\rangle_{\psi}^{2}+4)\gamma}{8})\Delta t+\frac{\sqrt{\gamma}\langle X\rangle_{\psi}}{2}\Delta W\right]\phi_{3},\\
   \Delta\phi_{4}=& (\frac{ \gamma\langle X\rangle_{\psi}}{2}\Delta t+  \sqrt{\gamma}\Delta W) (\phi_{2}+\phi_{3})  -\left[(\frac{i}{\hbar}H^{4,4}_{\text{eff}}+\frac{\gamma\langle X\rangle_{\psi}^{2}}{8})\Delta t+\frac{\sqrt{\gamma}\langle X\rangle_{\psi}}{2}\Delta W\right]\phi_{4}.
 \end{aligned}   
 \label{Eq:SSEcom}
\end{equation}
We find the corresponding Fokker-Planck equation of the component $\phi_{1}$ for the probability density $P(\phi_{1},t)$ is accordingly
\begin{equation}
\frac{\partial P(\phi_1, t)}{\partial t} = \frac{\partial}{\partial \phi_1} \left[ \left( \frac{i}{\hbar}H^{1,1}_{\text{eff}} + \frac{\langle X \rangle_\psi^2 + 8\gamma}{8} \right) \phi_1 P(\phi_1, t) \right] + \frac{1}{2} \frac{\partial^2}{\partial \phi_1^2} \left[ \left(\frac{\sqrt{\gamma} \langle X \rangle_\psi}{2} \phi_1 \right)^2 P(\phi_1, t) \right],
\end{equation}
where the fixed point can be identified by Eq.~(\ref{Eq:SSEcom}), and the steady-state solution is $P_{\text{ss}}(\phi_{1})=\delta(\phi_{1})$. 
This solution implies that the component $\phi_{1}$ will eventually vanish entirely in the long-time limit, and the wave function evolves within a three-level subspace. 
This is consistent with our previous discussion. 
The expectation value $\langle X\rangle_{\psi}$ will be zero to ensure the stochastic Wiener increment does not have an effect. 
In our work, the operator $X$ is $X=\sigma^{x}_{1} + \sigma^{x}_{2}$ for when $o=\sigma^{-}_{1} + \sigma^{-}_{2}$; and $X=\sigma^{x}_{1} - \sigma^{x}_{2}$ for when $o=\sigma^{-}_{1} - \sigma^{-}_{2}$. In this way, we prove that the oscillations in our system are robust and explain why the system spontaneously undergoes transitions in the type of synchronization under the action of different ancilla states.

\section{ Noise robustness of complete quench sequence}

In the main text, we show the density plot of the Pearson correlation evolution for the complete sequence as a function of collision number and noise strength [Fig.~3(a) of the main text].  
The synchronization transition remains evident regardless of the noise strength.
Here, we provide a detailed analysis of the complete quench sequence, comparing it with Figs.~3(c)-(f).

\begin{figure}
    \includegraphics[width=0.85\linewidth]{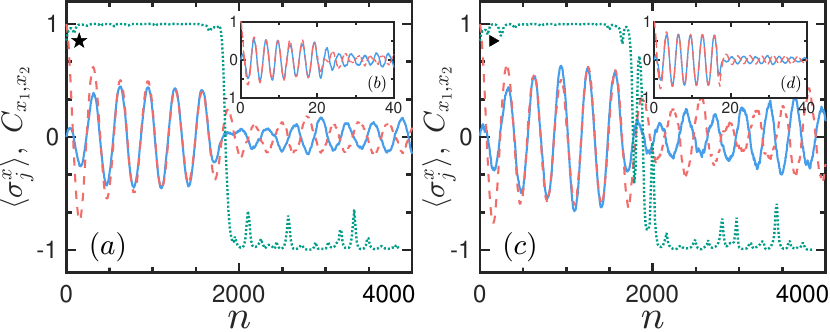}
    \caption{\label{Fig:NoiseEnd} The blue solid, red dashed, and green dotted lines show $\langle\sigma^{x}_{1}\rangle$, $\langle\sigma^{x}_{2}\rangle$, and $C_{x_{1},x_{2}}$ as a function of collision number $n$ under the complete quench sequence. The noise strengths are $\bar{\xi}=0.21$ (left) and $\bar{\xi}=0.49$ (right), respectively.} 
\end{figure}

Figures~\ref{Fig:NoiseEnd}(a) and (c) show the evolution of order parameters and Pearson correlation, represented by the blue solid, red dashed, and green dotted lines, respectively. 
In both cases, we observe a clear transition from in-phase to anti-phase synchronization, indicating the robustness to noise. 
Similarly, we observe synchronizations and the transition within single trajectories in both the weak [Fig.~\ref{Fig:NoiseEnd}(b)] and strong [Fig.~\ref{Fig:NoiseEnd}(d)] noise cases. 
These results demonstrate the system's capacity to adapt and maintain synchronization dynamics under varying noise strengths.

\section{Experimental considerations based on quantum circuits}

In this section, we discuss the scheme of utilizing quantum circuits as the experimental implementation platform and provide more details about the quantum algorithms.

As we mentioned in Sec.~\ref{SezcI}, in this collision model, the total unitary evolution operator for a complete collision process can be represented as $U_{\text{tot}} = U_{I}U_{S}$. 
The operator $U_{S}$ describes the free-evolution process and can be directly decomposed as 
\begin{equation}
\begin{aligned}
U_{S}(\tau)=&\exp[-i\tau\omega(\sigma^{z}_{1}+\sigma^{z}_{2})],\\
=&\exp[-i\tau\omega\sigma^{z}_{1}]\exp[-i\tau\omega\sigma^{z}_{2}],\\
=&U^{(1)}_{S}(\tau)U^{(2)}_{S}(\tau).
\end{aligned}
\end{equation}
This evolution operator acts locally on the two system qubits and can be conveniently represented using two $\texttt{U3}$ gates in Qiskit. 
The $\texttt{U3}$ gate is a single-qubit quantum gate that enables universal rotation about the Bloch sphere using three Euler angles.

\begin{figure}[!htpb]
    \centering
    \includegraphics[width=1\textwidth]{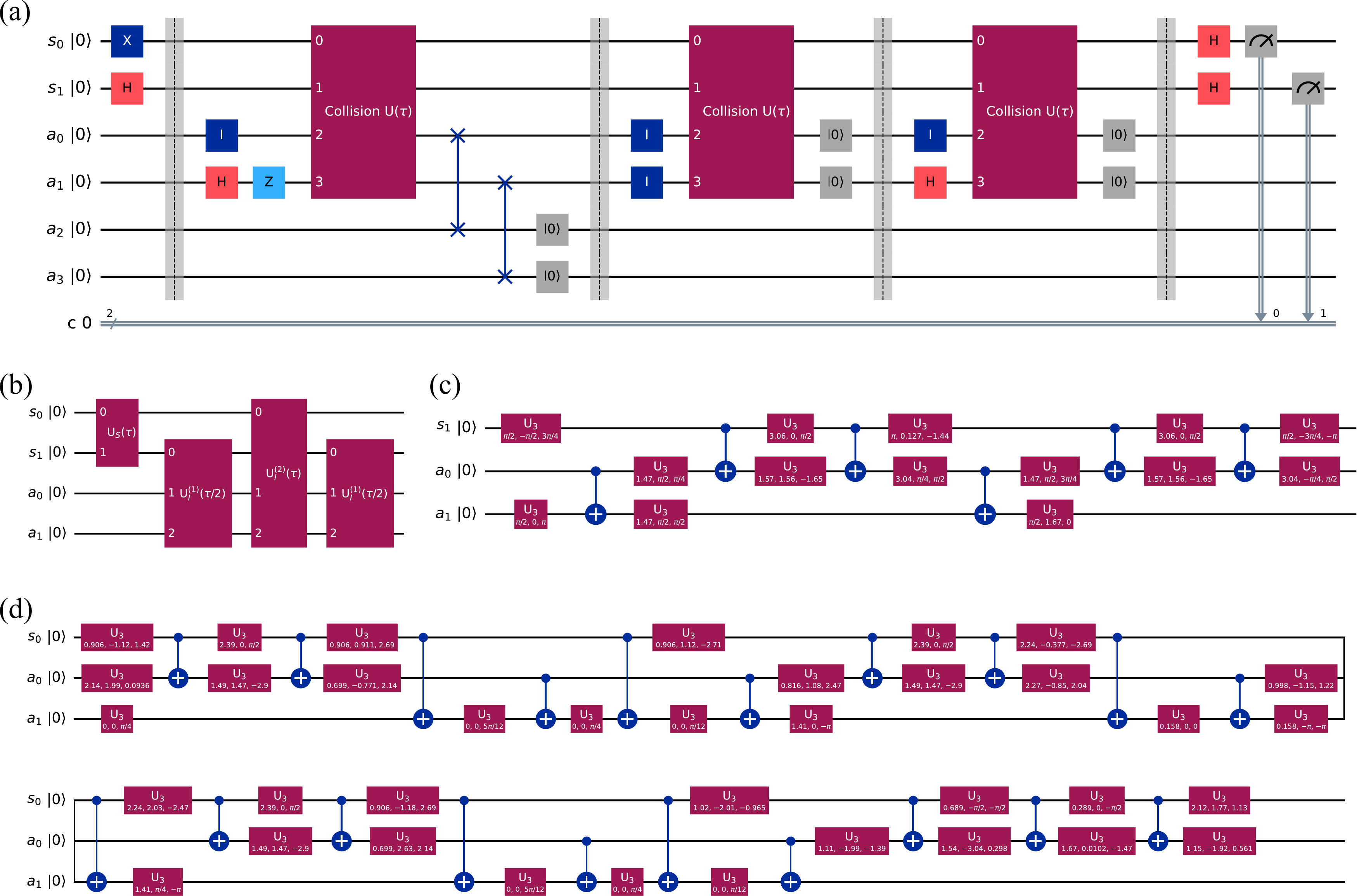} 
    \caption{\label{Fig:Circuit} The detailed decomposition of quantum circuits for the experimental implementation. (a) The experimental quantum circuit scheme consists of three main parts: the initial state preparation, unitary evolution process, and measurement stage. The unitary evolution part is illustrated with three collisions as an example, each of which corresponds to a different ancilla state. The circuit scheme involves six qubits, including two system qubits ($s_{0}$, $s_{1}$) and four ancilla qubits ($a_{0}$ to $a_{3}$). All qubits are initialized in state $|0\rangle$ [$|\downarrow\rangle$ in our QCM]. (b) Every single collision $U(\tau)$ consists of two parts: the free evolution of the system and the interaction between the system and ancillas. We perform the second-order Suzuki-Trotter decomposition of the second part, which is decomposed into a combination of $U^{(1)}_{I}(\tau)$ and $U^{(2)}_{I}(\tau/2)$. (c)-(d) The detailed decompositions of the gates $U^{(1)}_{I}(\tau)$ and $U^{(2)}_{I}(\tau/2)$, respectively, with $\tau=0.1$.}    
\end{figure}

In the scenario where the two qubits interact simultaneously with environment states, we treat the interaction evolution operator $U_{I}$ by the second-order Suzuki-Trotter decomposition, which is expressed as 
\begin{equation}
   U(\tau)=e^{-i\tau\sum_{k=1}^{L}H_{k}}=(\prod_{j=1}^{L} e^{-iH_{j}\tau/2m}\prod_{k=L}^{1} e^{-iH_{k}\tau/2m})^{m}+o(\tau^{3}/n^2).
   \label{SEq:Suzuki}
\end{equation}
We partition the interaction Hamiltonian into two terms $H_{I} = H_{I}^{(1)}+H_{I}^{(2)}$,  $H_{I}^{(1)} = g(\sigma^{-}_{1}|r\rangle\langle g|+\text{H.c.})$ and $H_{I}^{(2)} = g(\sigma^{-}_{2}|r\rangle\langle e|+\text{H.c.})$, and employ Eq.~(\ref{SEq:Suzuki}) with $m=1$ to derive the $U_{I}$ operator into
\begin{equation}
\begin{aligned}
U_{I}(\tau)&=\exp[-i\tau(H_{I}^{(1)}+H_{I}^{(2)})],\\
&\approx \exp(-iH_{I}^{(1)}\tau/2)\exp(-iH_{I}^{(2)}\tau)\exp(-iH_{I}^{(1)}\tau/2),\\
&=U^{(1)}_{I}(\tau/2)U^{(2)}_{I}(\tau)U^{(1)}_{I}(\tau/2),
\end{aligned}
\end{equation}
the total unitary evolution operator now is expressed as $U_{\text{tot}}=U_{I}^{(1)}(\tau/2)U_{I}^{(2)}(\tau)U_{I}^{(1)}(\tau/2)U^{(1)}_{S}(\tau)U^{(2)}_{S}(\tau)$.

As we mentioned in the main text, the environment (ancilla) is a qutrit and we map it into two qubits: $|g\rangle\mapsto|\downarrow\downarrow\rangle_{\text{QC}}$, $|e\rangle\mapsto|\downarrow\uparrow\rangle_{\text{QC}}$ and $|r\rangle\mapsto|\uparrow\downarrow\rangle_{\text{QC}}$. 
With the help of this mapping, the environment operators, i.e., $|r\rangle\langle g|$ and $|r\rangle\langle e|$ can be rewritten as $(\sigma^{x}\otimes\sigma^{z}+\sigma^{x}\otimes\mathbb{I}-i\sigma^{y}\otimes\sigma^{z}-i\sigma^{y}\otimes\mathbb{I})/4$ and $(\sigma^{x}\otimes\sigma^{x}+i\sigma^{x}\otimes\sigma^{y}-i\sigma^{y}\otimes\sigma^{x}+\sigma^{y}\otimes\sigma^{y})/4$, respectively. 
In this way, $H^{(1)}_{I}$ can be reexpressed as $(\sigma^{x}\otimes\sigma^{x}\otimes\sigma^{z}+\sigma^{x}\otimes\sigma^{x}\otimes\mathbb{I}-\sigma^{y}\otimes\sigma^{y}\otimes\sigma^{z}-\sigma^{y}\otimes\sigma^{y}\otimes\mathbb{I})/4$ and $H^{(2)}_{I}$ is $(\sigma^{x}\otimes\sigma^{x}\otimes\sigma^{x}+\sigma^{x}\otimes\sigma^{y}\otimes\sigma^{y}+\sigma^{y}\otimes\sigma^{x}\otimes\sigma^{y}-\sigma^{y}\otimes\sigma^{y}\otimes\sigma^{x})/4$.

In Fig.~\ref{Fig:Circuit}, we present the detailed decomposition of quantum circuits for our experimental implementation. 
Figure~\ref{Fig:Circuit}(a) specifically showcases the experimental quantum circuit scheme. 
The proposed scheme encompasses a total of three parts: the first part is the preparation of the system's initial state, which includes applying specific single-qubit gates, $\texttt{X}$ gate and $\texttt{H}$ gate on the system qubits $s_{0}$ and $s_{1}$ to prepare an initial stste $|\psi\rangle=\frac{|\uparrow\uparrow\rangle+|\downarrow\uparrow\rangle}{\sqrt{2}}$. 
The second part describes the quantum collision processes. Each completed collision process consists of three essential components: 
\begin{itemize}
    \item The preparation of ancilla states. The scheme of preparing the three different ancillas, which are required in the main text is depicted in Fig.~\ref{Fig:Circuit}(a) as an example. 
    \item The free-evolution part and system-ancilla interaction. After applying the second-order Suzuki-Trotter decomposition to the system-ancilla interaction part, this component contains four terms in order, as shown in Fig.~\ref{Fig:Circuit} (b). The free-evolution operator $U_{S}(\tau)$ can be realized by applying the $\texttt{U}_{3}$ gates, as we mentioned before. The system-ancilla interaction part is more complex. Here we set the collision duration to be $\tau=0.1$, and the interactions can be decomposed into a series of single-qubit ($\texttt{U3}$) and two-qubit ($\texttt{CNOT}$) gates. The specific forms of $U^{(1)}_{I}(\tau/2)$ and $U^{(2)}_{I}(\tau)$ are shown in Figs.~\ref{Fig:Circuit}(c) and (d), respectively.
    \item Refreshing the ancilla state. In the collision model setup, a sequence of ancillas is required to interact with the system. However, in a real quantum computer, direct interactions between the system qubits and arbitrary ancilla are not feasible due to the limitations imposed by the topology of the backend. Here we present two schemes for refreshing the ancilla states in Fig.~\ref{Fig:Circuit}(a): (i). Refreshing the states of $a_{0}$ and $a_{1}$ by exchanging the latter with the remaining ancillas of the train through $\texttt{SWAP}$ gates, and this scheme requires a sufficient number of ancilla qubits. (ii). Reinitializing the states of $a_{0}$ and $a_{1}$ into $|0\rangle$ incoherently by the $\texttt{reset}$ gates.
\end{itemize}
The final step is to measure the system qubits. By applying the $\texttt{H}$ gates to transform the basis into the Pauli $x$-direction, then we can directly obtain the expectation values of $\langle \sigma^{x}_{j}\rangle$ by the classical measure.

\bibliography{Ref}